\documentclass[12pt]{article}

\usepackage{amsmath,amssymb,amsthm}

\newcommand\labfig[1] {\label{fig:#1}}

\newcommand{\beqa}{\begin{eqnarray}}
\newcommand{\eeqa}[1]{\label{#1}\end{eqnarray}}
\newcommand{\beq}{\begin{equation}}
\newcommand{\eeq}[1]{\label{#1}\end{equation}}

\newcommand{\Gt}{\theta}

\def \ba {\begin{array}}
\def \ea {\end{array}}

\def \refe #1.{(\ref{#1})}
\def \reff #1.{figure~\ref{#1}}
\def \refs #1.{section~\ref{#1}}
\def \refss #1.{subsection~\ref{#1}}
\def \refD #1.{Definition~\ref{#1}}
\def \refT #1.{Theorem~\ref{#1}}
\def \refL #1.{Lemma~\ref{#1}}
\def \refC #1.{Corollary~\ref{#1}}
\def \refP #1.{Proposition~\ref{#1}}
\def \refR #1.{Remark~\ref{#1}}
\def \refE #1.{Example~\ref{#1}}
\def \refN #1.{Notation~\ref{#1}}

\usepackage{graphics}
\usepackage{amssymb}
\begin{document}
\vspace{-1in}
\title{New examples of three-dimensional dilational materials}
\author{Graeme Walter Milton\\
\small{Department of Mathematics, University of Utah, Salt Lake City UT 84112, USA}}
\date{}
\maketitle
\begin{abstract}
Two-dimensional dilational materials, for which the only easy mode of deformation is a dilation are
reviewed and connections are drawn between models previously proposed in the literature. Some models
which appear to be dilational materials, but which in fact are not, are also discussed. Finally, four new
examples of three-dimensional dilational materials are given.
\end{abstract}
\vskip2mm
\noindent Keywords: auxetics, dilational material, negative Poisson's ratio  
\noindent 
\vskip2mm

\section{Introduction}

Auxetic materials are (possibly anisotropic) materials with negative Poisson's ratios and have attracted considerable
attention: see the review of \cite{Greaves:2011:PRM}, the papers in this volume of Physica Status Solidi B, and references therein.
A dilational material is an auxetic material for which a dilation is the only easy mode of deformation under finite deformations,
and as a consequence they prefer to keep their shape (but not size) as they are deformed. Ideal dilational materials are rigid against
all other deformations, or if flexible to other deformations then dilations cost no energy. This requires either sliding surfaces or ideal hinge type junctions,
which however can approximated. In materials that are approximations
to the ideal, dilations cost little energy by comparison to deformations that are not close to dilations. Ideal dilational materials, if 
flexible, are not necessarily elastically isotropic, but they have a Poisson's ratio of $-1$ in all directions over a range of strains. For
example under infinitesimal deformations such a material with cubic symmetry could have a bulk modulus which is zero but two different shear
moduli (the idealization of the material constructed by B{\"{u}}ckmann et.al. \cite{Buckmann:2014:TDD} falls into this category). 

The first example of a two or three dimensional dilational material
appears to be the telescoping rod model of  Rothenburg, Berlin and Bathurst \cite{Rothenburg:1991:MIM}, the simplest realization of which is 
shown in figure 1a, (see also 
figure 3 in \cite{Milton:1992:CMP} and figure 3a in \cite{Lakes:2008:NCN}). The telescoping rods have sliding surfaces, and so in any realization of this material
there would be frictional forces and the possibilty of sticking. If the microstructure is scaled to a very small size, and the telescoping rods reduced in proportion,
then the total frictional surface area per unit volume will become very large, and friction will be an important consideration. Alternatively, in three dimensions
(but not in two dimensions) the frictional area can be reduced by scaling the telescope diameter to be much smaller than its length, as its length is reduced.
However, in this case, bending of the telescopes will have to be taken into account. 
An early model, with sliding surfaces, having a Poisson's ratio of $-1$ for infinitesimal deformations, was constructed by Almgren \cite{Almgren:1985:ITD}, but it is not
a dilational material as it has an anisotropic response at large deformations. A dilational material without sliding surfaces was 
constructed by Milton \cite{Milton:1992:CMP}, see figure 5 in that paper, and appears to be the first model with a negative Poisson's ratio having chirality.
Related to this chiral linkage model (see figure 2)
is an elegant structure (composed of rotating hexagons and triangles) discovered by Mitschke et.al. \cite{Mitschke:2011:FAF}: see also \cite{Mitschke:2013:SDA}.

\begin{figure}[htbp]
\vspace{4.0in}
\hspace{0.0in}
{\resizebox{1.0in}{0.5in}
{\includegraphics[0in,0in][4in,2in]{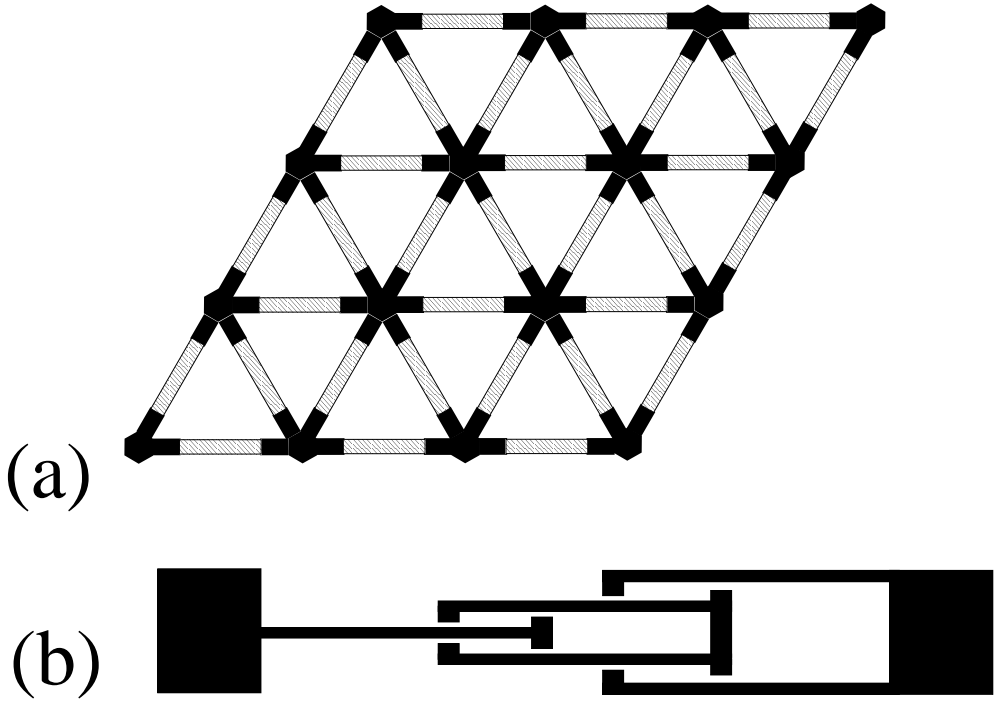}}}
\vspace{0.1in}
\caption{An hexagonal array of telescoping rods welded at fixed angles of $60^\circ$ at the junctions, as shown in (a), is one of the simplest realizations of the 
dilational material first suggested in \protect\cite{Rothenburg:1991:MIM}. As recognized in that paper such telescoping rods are easily 
realized if a sliding joint is used in the middle of each rod. Even in two-dimensions, the rods can expand to an 
arbitarily large distance and not fall apart if one allows for multiple
sheaths in the telescoping rod as illustrated in (b): for clarity the sliding surfaces are shown with small gaps. 
In this and in subsequent figures, except for figure 4, everything that is black is rigid or at least
comparatively very stiff.}
\labfig{1}
\end{figure}

\begin{figure}[htbp]
\vspace{1.5in}
\hspace{0.0in}
{\resizebox{1.0in}{0.5in}
{\includegraphics[0in,0in][10in,5in]{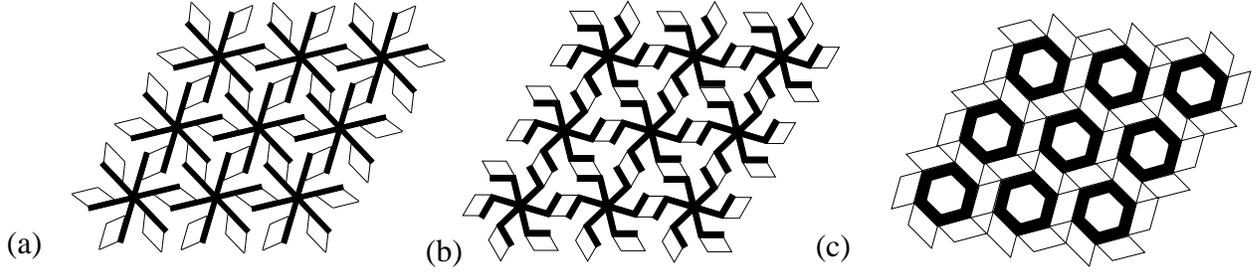}}}
\vspace{0.1in}
\caption{If one wants to avoid sliding surfaces, one can instead use the chiral linkage model (a)
proposed in \protect\cite{Milton:1992:CMP}, where the point contacts between the rigid black spoked cores
and the rods are treated as hinges. 
By bending the spokes in the cores one transitions from (a) to (b) and 
finally when the spokes are bent at an angle of $120^\circ$ to the structure (c) proposed by Mitschke et.al. \protect\cite{Mitschke:2011:FAF}}
\labfig{2}
\end{figure}

 Another especially simple dilational material is the rotating
squares model of Grima and Evans \cite{Grima:2000:ABR} (see figure 3) that is a simplification of an earlier 
model of Sigmund \cite{Sigmund:1995:TMP} (see his figure 4) which was recently rediscovered and generalized
by Cabras and Brun \cite{Cabras:2014:ATD}. Sigmund \cite{Sigmund:1995:TMP} obtained dilational materials
built from rotatable frames in both two and three dimensions, but these have sliding surfaces. 
Dilational materials can also exhibit arbitarily large flexibility
windows even without using sliding surfaces (such as telescoping rods with multiple sheaths as in figure 1b): see the
unit cell in figure 8 in \cite{Milton:2013:CCM}. A three-dimensional rotating cuboid structure was constructed by 
Attard and Grima \cite{Attard:2012:TDR}, which exhibits auxetic but not dilational behavior. 

\begin{figure}[htbp]
\vspace{2.0in}
\hspace{1.0in}
{\resizebox{1.0in}{0.5in}
{\includegraphics[0in,0in][5in,2.5in]{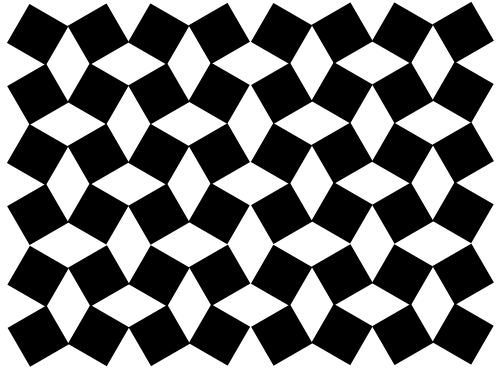}}}
\vspace{0.1in}
\caption{The rotating array of squares of Grima and Evans \protect\cite{Grima:2000:ABR} is another extremely simple dilational material.}
\labfig{3}
\end{figure}

Recently, three dimensional dilational materials without sliding surfaces have been discovered (see the unit cell in figure 19b of \cite{Milton:2013:CCM},
and \cite{Buckmann:2014:TDD}). In particular the model in \cite{Buckmann:2014:TDD} was investigated in detail, and moreover 
experimentally realized and tested, exhibiting a Poisson's ratio as low as $-0.8$. Previously polymer foams had been found to
have Poisson's ratio values below $-0.6$ maintained over about $20\%$ strain \cite{Choi:1992:NPP}, and in metal foams
a Poisson's ratio of $0.8$ had been achieved \cite{Choi:1992:NPM}, although it is not clear if these non-periodic foam structures would permit
a Poisson's ratio approaching $-1$, if suitably manufactured.  

This paper develops four additional examples of three-dimensional dilational materials without sliding surfaces. 
The ideal dilational materials discussed here which do not have sliding surfaces incorporate hinges. These hinges can be appoximated in various ways
without the use of sliding surfaces as illustrated in figure 4. The models presented here show the variety of ways dilational materials can be designed, 
and could serve as blueprints for the physical construction of realistic materials which are close to being dilational. 

\begin{figure}[htbp]
\vspace{3.0in}
\hspace{1.0in}
{\resizebox{1.0in}{0.5in}
{\includegraphics[0in,0in][2in,1.0in]{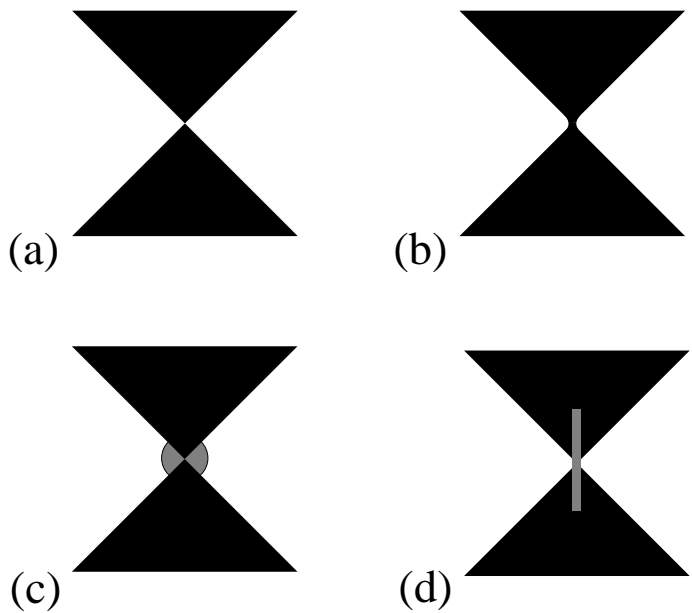}}}
\vspace{0.1in}
\caption{An ideal hinge as in (a) of infinitely rigid material surrounded by void can be approximated by very stiff material with a thin neck as in (b), as suggested
for example in \protect\cite{Milton:1992:CMP,Kadic:2012:PPM}. Alternatively the junction of very stiff material, may be clad by softer rubber represented by the shaded region in (c). A better alternative as in (d) may be to glue into the very stiff material a fiber(or fibers) which is (or are) difficult to stretch but easy to bend,
to minimize the axial compliance.}
\labfig{4}
\end{figure}

\section{Models which are not dilational materials}

Some models which at first sight appear to be dilational materials, turn out not to be upon closer investigation. An example
is the chiral honeycomb of Prall and Lakes \cite{Prall:1996:PCH} as shown in figure 5a. As recognised by them, deformation of 
this honeycomb under finite dilation requires some energy, due to the bending of the ribs as in figure 5b. Suppose we have
a beam of such material, $M$ cells wide and $L$ cells long which we clamp in the middle between two lubricated rectangular blocks as in
figure 5c. The question is whether the deformation will be one of uniform dilation, costing an energy $E_0$ per unit cell
that is a total energy of $W_0=MLE_0$, or whether the deformation will be closer to that illustrated in figure 5c. To show that it
is not a uniform deformation it suffices to obtain a trial deformation field with lower energy. To do so we choose
a trial field where the deformation is zero outside a transition region, and inside the transition region which is $N$
cells wide is allowed to deform in some way (independent of $L$) compatible with the boundary conditions at the clamps. Let
$W$ be the total energy of the trial deformation, which is independent of $L$. Then if $L$ is such that $W_0=MLE_0>W$ the
prefered deformation will not be one of uniform dilation. Note that as the thickness of ribs approaches zero, while the moduli
of the ribs are appropriately rescaled, the energies $E_0$ and $W$ will approach fixed limits: thus the prefered deformation
will not be one of uniform dilation even in the limit as the ribs are infinitesimally thin. The crucial point is that the width of the 
actual transition region remains almost independent of $L$ and almost independent of the thinness of the ribs, for sufficiently large $L$
and for sufficiently thin ribs. By contrast if we did the same experiment with approximations to the dilational materials discussed here
(with hinges of the type described in figures 4(b), 4(c) and 4(d)) the width of the actual transition region would continue to increase as the approximation 
to the ideal material improves so that ultimately the beam would undergo approximately uniform dilation for fixed $L$ no matter how large $L$ is. 
The difference is that in the approximations to the dilational material, the energy required to splay (taper) the material is enormously larger than
the energy to uniformly compress it, with the ratio approaching infinity as the material becomes more ideal. This is not true of the
chiral honeycomb: when the chiral honeycomb is splayed it is only necessary that the ribs be bent a little more than under compression. Nevertheless, it may be the case that the chiral honeycomb model can be modified to obtain a dilational material (perhaps by tapering the endpoints of the ribs down to points meeting the circles at an angle, rather than tangentially, and appropriately rescaling the moduli.)

\begin{figure}[htbp]
\vspace{4.0in}
\hspace{0.0in}
{\resizebox{1.0in}{0.5in}
{\includegraphics[0in,0in][7in,3.5in]{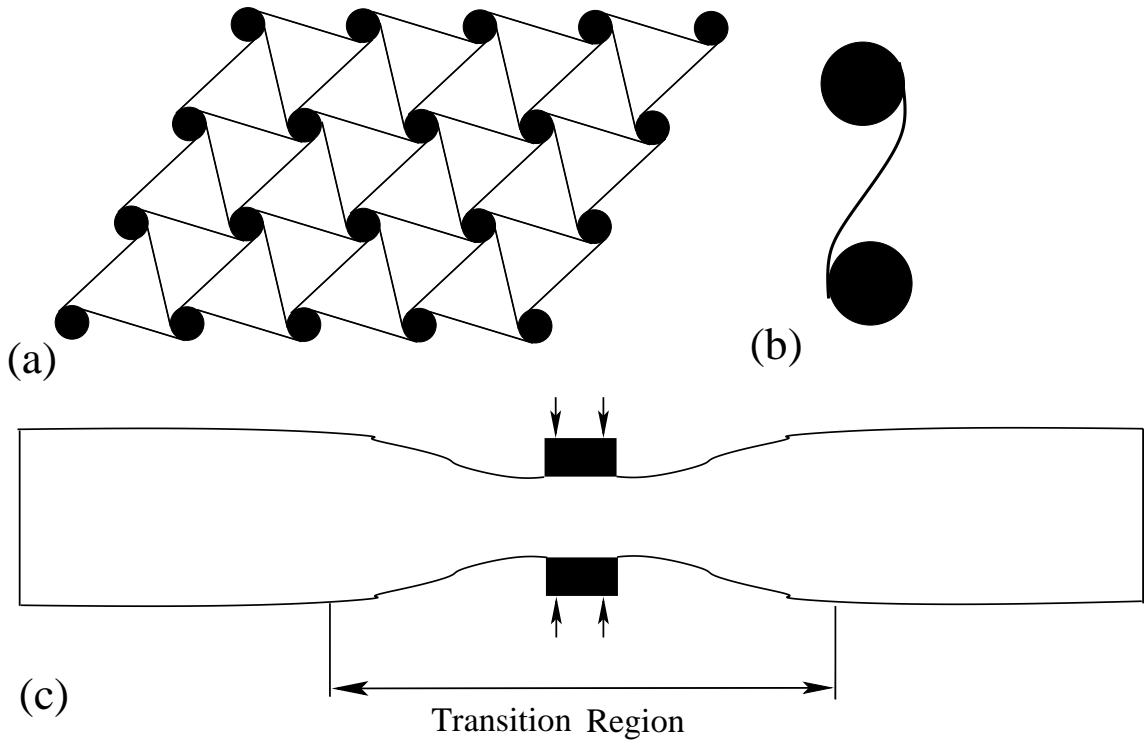}}}
\vspace{0.1in}
\caption{The chiral honeycomb of Prall and Lakes \protect\cite{Prall:1996:PCH}, as illustrated in (a), has for
infinitesimal deformations a Poisson's ration of $-1$ in the limit as the flexible ribs joining the circles become very thin. However it is not a dilational material. The energy stored in the flexure of each
rib, bent as in (b) under finite compression, causes a long beam of the material to approximately 
deform as in (c) when the middle is clamped between two lubricated rectangles, and the length of the transition region remains much less than $L$, for sufficiently large
$L$ as the ribs 
become increasingly thinner.}
\labfig{5}
\end{figure}

Another model which is not a dilational material is the two-dimensional variant of the three-dimensional model proposed in \cite{Buckmann:2014:TDD}. 
This variant illustrated in figure 6a, has macroscopic easy deformations which are non-affine. To understand this, take one row of the model as in figure
6b, and consider its deformations as in figure 6c. Within limited ranges the position of the points $P_i$, $i=1,2,3,\ldots$ can be chosen independently
as can be the angles $\Gt_i$, $i=1,2,3,\ldots$. Taken together they (and the geometry) determine the position of the  points $Q_i$, $i=1,2,3,\ldots$ at
the top of this row. The next row can be joined to these points, and so on, leaving a lot of internal parameters (the angles $\Gt_i$, $i=1,2,3,\ldots$ in each row)
that can be freely varied (within limits), and which (along with the positions of the points $P_i$, $i=1,2,3,\ldots$ of the first row) 
determine the possible macroscopic deformations of the structure. (It may be necessary to assume the structure has bounded extent). This argument
does not apply to the three dimensional dilational material proposed in \cite{Buckmann:2014:TDD} since the orthogonality of the panels in that
construction guarantees that the macroscopic easy deformation will be affine, and necessarily a dilation.

\begin{figure}[htbp]
\vspace{3.0in}
\hspace{0.0in}
{\resizebox{1.0in}{0.5in}
{\includegraphics[0in,0in][3.5in,1.75in]{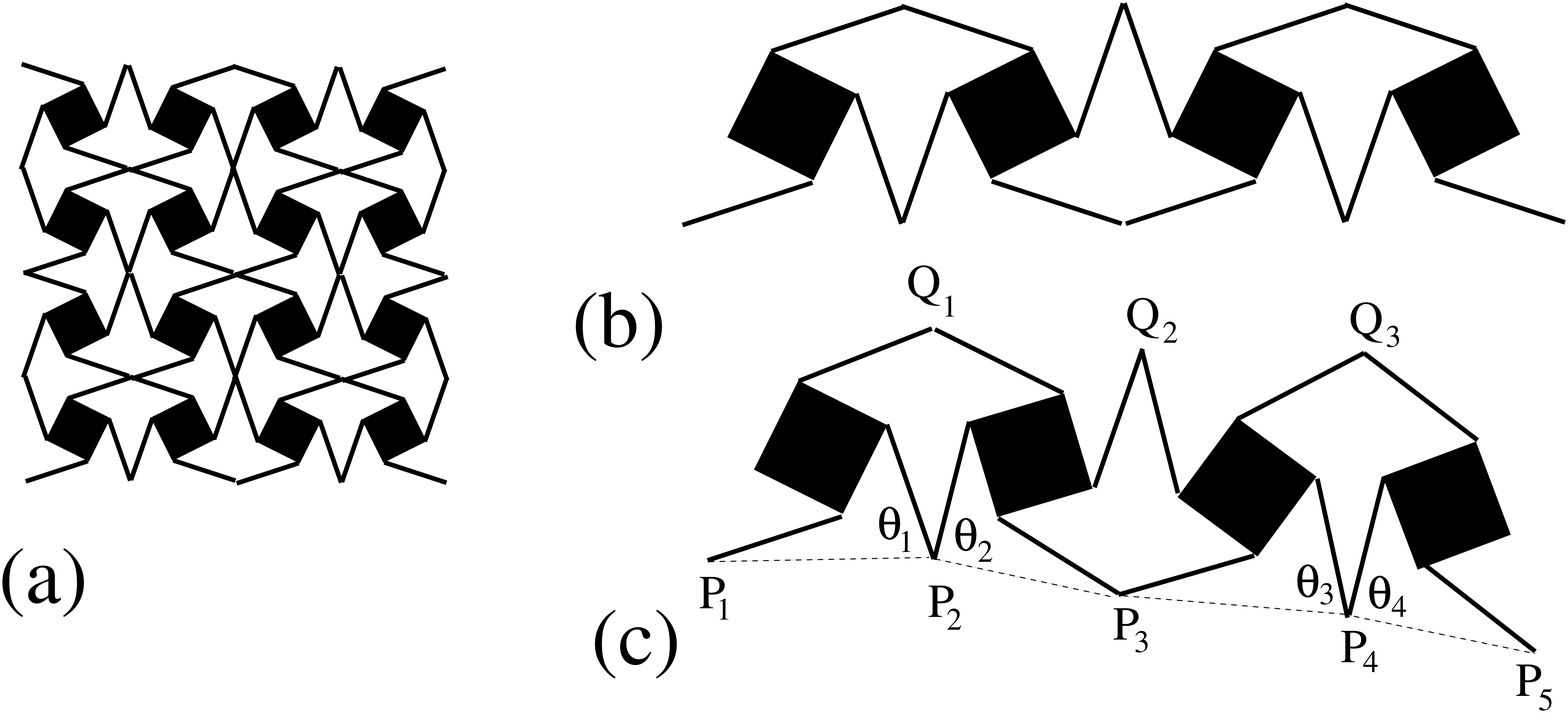}}}
\vspace{0.1in}
\caption{Shown in (a) is the two-dimensional variant of the three-dimensional model proposed in \protect\cite{Buckmann:2014:TDD}. One row of the structure
as in (b) can deform as in (c), and additional rows built on top of this one can deform similarly. See the text for more details.}
\labfig{6}
\end{figure}

\section{New three-dimensional dilational materials}

Rothenburg, Berlin and Bathurst \cite{Rothenburg:1991:MIM} recognized that their telescoping rod model, our figure 1, extends to three dimensions. They
suggested taking a random, statistically homogeneous and isotropic network of trusses and replacing the trusses by telescoping rods bolted at fixed angles at the
junctions. Alternatively one could put telescoping rods replacing the nearest
neighbour bonds in a face centered cubic lattice, as suggested by Lakes and Wojciechowski \cite{Lakes:2008:NCN}, or in a hexagonal close packed lattice, 
with the ends of the rods welded at the vertices. Replacing the nearest neighbour bonds by telescoping rods in some other lattices such as the cubic or hexagonal lattice 
will not work as these lattices allow some easy modes of deformations in addition to dilations. If one wishes to avoid sliding surfaces 
in each telescoping rod, one way of doing this is to use rods which have Sarrus linkages (figure 7) in their center. This is an especially easy way to obtain three-dimensional dilational materials. Interestingly the Sarrus linkage appears in the rotating cuboid structure of Attard and Grima \cite{Attard:2012:TDR},  which exhibits auxetic (but not dilational) properties.

\begin{figure}[htbp]
\vspace{1.0in}
\hspace{1.0in}
{\resizebox{1.0in}{0.5in}
{\includegraphics[0in,0in][4in,2in]{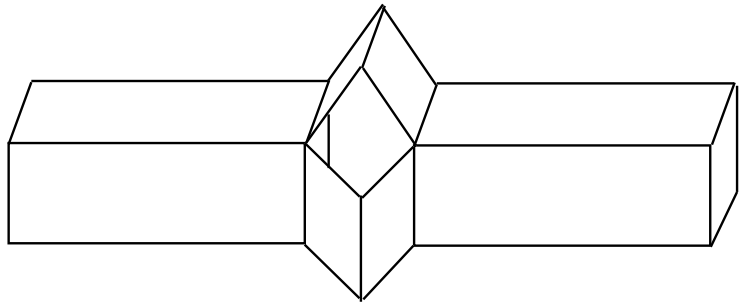}}}
\vspace{0.1in}
\caption{In three dimensions a Sarrus linkage is a way of creating a telescoping beam without using sliding
surfaces.}
\labfig{7}
\end{figure}

The three-dimensional rotating squares model of Sigmund \cite{Sigmund:1995:TMP} is easily modified to avoid sliding surfaces if the rotating square plates on each face
of the unit cube are curved (with positive curvature) so they no longer slide against each other, as sketched in figure 8.

\begin{figure}[htbp]
\vspace{2.0in}
\hspace{1.0in}
{\resizebox{1.0in}{0.5in}
{\includegraphics[0in,0in][3in,1.5in]{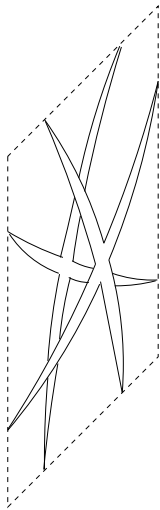}}}
\vspace{0.1in}
\caption{It is straightforward to modify the three dimensional rotating square model of Sigmund \protect\cite{Sigmund:1995:TMP} to obtain a dilational material without
sliding surfaces. Here the 
rotating square plates are replaced by star shaped objects, whose arms are curved so that they do not slide against the other. The stars have some thickness, not shown 
here, but taper to points at the star tips. This structure tiles each square face in a cubic 
lattice: the boundary of the square face being  marked by the dashed line here.}
\labfig{8}
\end{figure}

The rotating squares model of Grima and Evans \cite{Grima:2000:ABR} (see our figure 3) can also be used as a basis for constructing three-dimensional dilational materials.
as illustrated in figure 9. In any horizontal plane going through a vertex in the model of figure 9c the cross sectional structure will look exactly like
the rotating squares model, and thus the macroscopic deformation in this plane must be one of uniform dilation. The same can be said
for the cross sectional structure in any plane at $60^\circ$ or $120^\circ$  going through a vertex in the model of figure 9c. Since the points $P$ are shared by the
planes the same dilational factor must be common to each plane. Thus the three-dimensional macroscopic deformation must be one of uniform dilation. 
\begin{figure}[htbp]
\vspace{4.0in}
\hspace{0.0in}
{\resizebox{1.0in}{0.5in}
{\includegraphics[0in,0in][3in,1.5in]{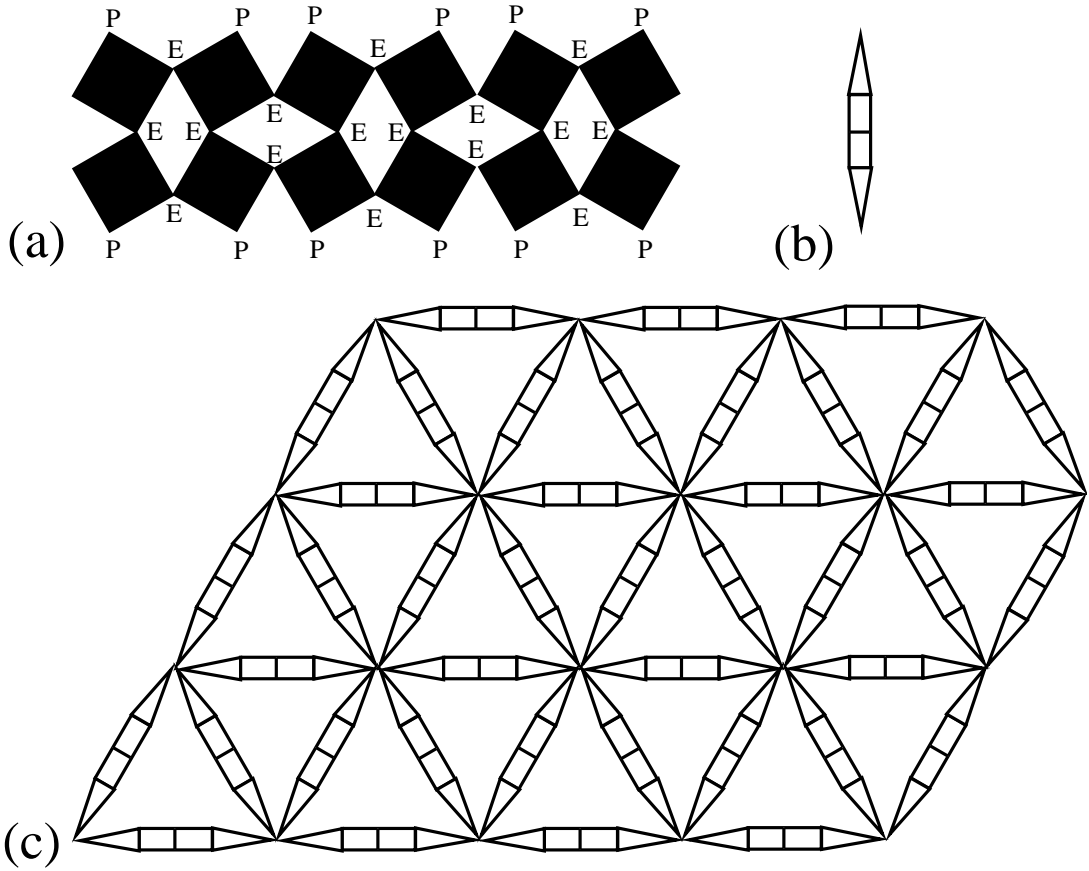}}}
\vspace{0.1in}
\caption{To construct a three-dimensional dilational material using the rotating squares model of Grima and Evans \protect\cite{Grima:2000:ABR} one constructs panels
which from the front look like (a) and from the side look like (b). The points $E$ denote edge hinges, and each panel is tapered at the points $P$. The
panels are then assembled into a hexagonal lattice, meeting at the points $P$ at the vertices of the lattice.}
\labfig{9}
\end{figure}

Motivated by the three-dimensional dilational material, with the cubic unit cell in figure 19b of \cite{Milton:2013:CCM}, which has as its faces square
panels the vertices of which remain square as the material deforms, one can similarly consider equilateral triangular panels as in figure 10, which remain 
equilateral as the panel deforms. These panels could be joined together to form regular octohedra (8 panels for each octahedron), and the octohedra
stacked with tetrahedral cavities in-between to form a tetrahedral-octahedral honeycomb, which will then be yet another three-dimensional dilational material. 

\begin{figure}[htbp]
\vspace{2.5in}
\hspace{0.0in}
{\resizebox{1.0in}{0.5in}
{\includegraphics[0in,0in][9in,4.5in]{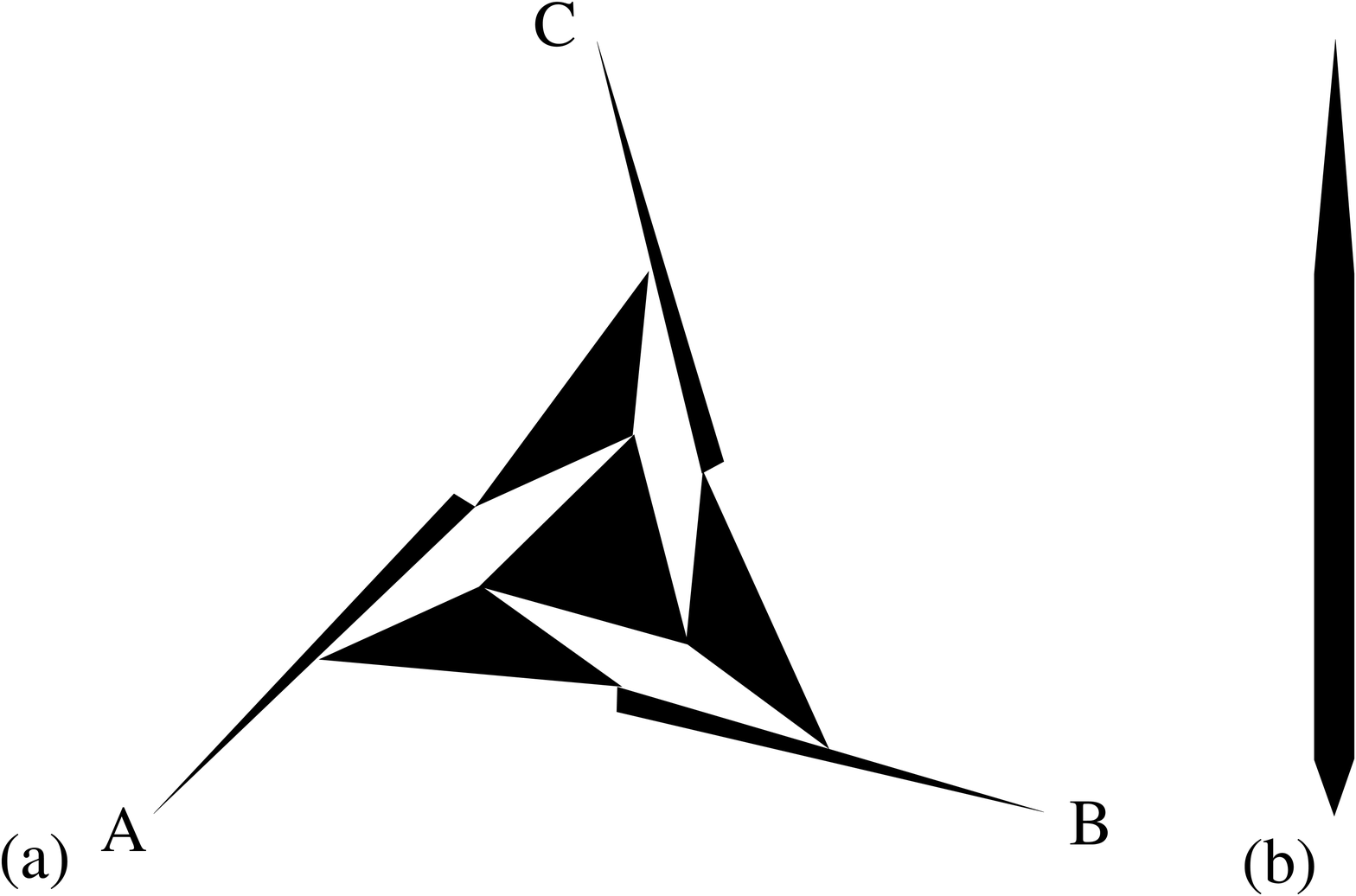}}}
\vspace{0.1in}
\caption{A triangular panel, with vertices $A$, $B$ and $C$ that always form an equilateral triangle as the structure is deformed. All interior junctions
are edge hinges, and the structure tapers to a point at the points $A$, $B$ and $C$. Figure (a) shows the panel from the front, and figure (b) the panel when viewed from the side.}
\labfig{10}
\end{figure}

\section*{Acknowledgements}
GWM is grateful to Martin Wegener and his group for stimulating this research and to the referees and Roderic Lakes for helpful comments on the manuscript. Additionally
GWM is thankful for support from the National Science Foundation through grant DMS-1211359.

\bibliographystyle{siam}
\bibliography{/home/milton/tcbook,/home/milton/newref}
\end{document}